# Fully Homomorphic Encryption Scheme with Symmetric Keys

*A*

*Dissertation*

*submitted*

*in partial fulfillment*

*for the award of the Degree of*

*Master of Technology*

*in* **Department of Computer Science & Engineering**

*(with specialization in Computer Engineering)*

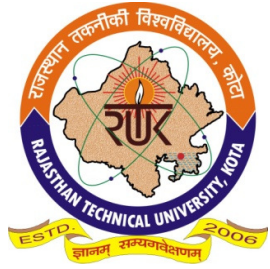

| Supervisor: | Submitted by: |
|---|---|
| C. P. Gupta | Iti Sharma |
| Asso. Prof & Head, CSE | Enrollment No.: 11E2UCCSF4XP606 |

**Department of Computer Science & Engineering**

University College of Engineering,

Rajasthan Technical University, Kota

**August – 2013**

# CANDIDATE'S DECLARATION

I hereby declare that the work, which is being presented in the Dissertation, entitled **"Fully Homomorphic Encryption Scheme with Symmetric Keys"** in partial fulfillment for the award of Degree of "Master of Technology" in Department of Computer Science and Engineering with Specialization in **Computer Engineering** and submitted to the **Department of Computer Science and Engineering,** University College of Engineering, Rajasthan Technical University is a record of my own investigations carried under the Guidance of  C. P. Gupta, Asso Prof & Head, Department of Computer Science and Engineering, UCE, Kota.

I have not submitted the matter presented in this Dissertation anywhere for the award of any other Degree.

Iti Sharma
Computer Engineering

Enrollment No.: 11E2UCCSF4XP606

University College of Engineering, RTU, Kota.

**Counter Signed By**

………………………..

C. P. Gupta,
Asso. Prof. and Head, CSE Dept

# CERTIFICATE

This is to certify that this dissertation entitled "**Fully Homomorphic Encryption with Symmetric Keys**" has been successfully carried out by **Iti Sharma (Enrollment No:11E2UCCSF4XP606),** under my supervision and guidance, in partial fulfillment of the requirement for the award of **Master of Technology** Degree in **Computer Engineering** from **University College of Engineering, Rajasthan Technical University, Kota** for the year **2011-2013.**

Place: Kota                                                                                           **Supervisor**

Date:                                                                                                        C. P. Gupta

                                                                                              Asso. Prof. & Head, CSE Dept

# Acknowledgment

It is my great privilege to express sincere gratitude & thanks to my supervisor **C. P. Gupta**, Associate Professor & Head, Department of Computer Science & Engineering, UCE, RTU, Kota for his valuable guidance during each and every phase of this work. The keen observation and motivating style of critique and compliment kept me stimulated towards perfection. I thank him for the interest and energy that was committed to the dissertation and for allowing me a wide academic freedom.

I would also thank my family for their help, support and patience. Not in the least I would thank Almighty for blessing me.

Iti Sharma
Computer Science Engineering
Enrollment No.: 11E2UCCSF4XP606
University College of Engineering,
RTU, Kota(Raj)

# List of Symbols and Notations

| | |
|---|---|
| $\mathbb{Z}_q$ | finite field of integers of order $q$ |
| $M_n(\mathbb{Z}_q)$ | square matrix of dimension n with elements from $\mathbb{Z}_q$ |
| $\parallel$ | concatenation of bit strings |
| $\mid$ | concatenation of matrices of equal number of rows, column-wise |
| $a \xleftarrow{\$} A$ | randomly select an item $a$ from set $A$ |
| $a_{1\ldots n} \xleftarrow{\$,n} A$ | randomly select n items from set $A$ |
| $A^{-1}$ | inverse of matrix A |
| A.B | multiplication of two matrices A and B |
| Enc(x,k) | Encryption of plaintext x under key k |
| Dec(c,k) | Decryption of ciphertext c under key k |
| $\tilde{O}(\lambda)$ | Landau notation equivalent to Asymptotic $O(\lambda \log \lambda \log \log \lambda)$ |

# List of Figures

# List of Tables

# List of Publications

1. Iti Sharma and Mayank Sharma, "Delegating Computations: Potentials of Homomorphic Encryption", 17th Annual Conference of GAMS and National Symposium on Computational Mathematics and Information Technology, JUET, Guna, Dec 2012 .

2. Iti Sharma and Heena Jain, "Cloud Integrity: Scope of Homomorphic Encryption", National Conference on Advancements in Information, Computer and Communication AICC-2013, Kautilya Institute of Technology and Engineering, Jaipur, March 2013.

3. C.P. Gupta and Iti Sharma, "A Fully Homomorphic Encryption scheme with Symmetric Keys with Application to Private Data Processing in Clouds", Fourth International Conference on Network of the Future (NoF'13), Pohang, Korea. [Accepted]

# CONTENTS

# Abstract

Homomorphic encryption has largely been studied in context of public key cryptosystems. But there are applications which inherently would require symmetric keys. We propose a symmetric key encryption scheme with fully homomorphic evaluation capabilities. The operations are matrix based, that is the scheme consists of mapping the operations on integers to operations on matrix. We aim at proposing an idea how a fully homomorphic scheme with symmetric keys can be inculcated into application like private data processing. We propose ideas for primitives required in a FHE scheme to make it practical and more useful. Certain applications which can benefit from homomorphic encryption involve more than one party, such as multiparty computation. Majority of the proposed schemes have not explored this area. Our proposal aims at answering this. The proposed scheme is computationally light, efficient, multi-hop, circuit-private and can be deployed in multiple user environment. It derives its security from hardness of factorizing a large integer, which is basis of many public key cryptosystems. Besides the primitives for encryption, decryption and evaluation, we have included primitives which are useful to adapt the scheme to specific applications of delegating computation and data access control in multiuser environments like that of cloud computing. We also include a protocol which uses the proposed scheme for private data processing in clouds. It can easily be extended for keyword search in indices of encrypted databases, PIR and electronic voting.

We also propose possible variants which give an idea of how can a fully homomorphic encryption scheme be designed using symmetric keys. We have also included a checklist of properties of a homomorphic scheme when employing it to certain application and tallied our proposal against it. We have presented security analysis of our scheme along with formal proofs. The performance of the scheme has been compared with current efficient schemes.

**Chapter 1**

# INTRODUCTION

The amount of data generated, stored and communicated electronically is growing exponentially year by year, and the related growth is the vulnerability of data hence the demand of making it secure. Cryptography has emerged as most effective data protection solution. At present, cryptographic primitives have provided both the data owners and users efficient means to ensure security of their data and algorithms in terms of confidentiality, integrity, authentication, validation, and verification.

## 1.1 Conventional Cryptography:

All cryptographic techniques in use today can be broadly classified as Symmetric and Asymmetric encryption.

- Symmetric or the secret key based cryptography implies using same key for both encryption and decryption. It is a cryptosystem defined by two algorithms. During communication, the sender uses the encryption algorithm Enc(m,k) where m is message to be encrypted and k is the secret key, to obtain a ciphertext c corresponding to plaintext m. This encrypted message is sent to the receiver. The receiver retrieves the message using decryption algorithm Dec(c,k).
- Asymmetric or public key based systems refer to use of different keys for encryption and decryption. The key known only to a sender (or receiver) is called the private key, and the key which is published and thus known to more than one party is called the public key. Encryption algorithm, Enc(m,pk) encrypts message m under public key pk, and the message is retrieved using decryption algorithm Dec(m,sk), where sk is the private key. In certain applications like digital signatures, encryption is performed using private key, hence decryption is done using the public key.

## 1.2 New Challenges Posed by Cloud Computing

The emergence of cloud computing where critical customer and enterprise data could be held by third party cloud providers in a public and/or shared (multi-tenant) computing and storage environments highlights the need to use encryption as a primary security control. Security threats and application of encryption mechanisms are discussed in context of "data at rest",

"data in transit", and "data in use". While the security of data in transit benefits from mature encryption tools such as SSL, protecting data at rest while ensuring its availability presents additional and ongoing challenges. Encryption of a database should not adversely affect the ability of applications to use this data. Hence, in a cloud computing scenario, encryption solutions must be architectured to achieve the goals of both data protection (confidentiality and integrity) as well as availability of the data, the service, and the capability to collaborate and share data easily. Neither symmetric nor asymmetric encryption methods completely suffice the needs of cloud computing environment. Here homomorphic cryptography comes into picture.

## 1.3 Homomorphic Cryptography

The aim of homomorphic cryptography is to ensure privacy of data in communication, storage or in use by processes with mechanisms similar to conventional cryptography, but with added capabilities of computing over encrypted data, searching an encrypted data, etc. Homomorphism is a property by which a problem in one algebraic system can be converted to a problem in another algebraic system, be solved and the solution later can also be translated back effectively. Thus, homomorphism makes secure delegation of computation to a third party possible. Many conventional encryption schemes possess either multiplicative or additive homomorphic property and are currently in use for respective applications. Yet, a fully homomorphic encryption scheme which could perform any arbitrary computation over encrypted data appeared in 2009 as Gentry's work [1].

## 1.4 Open problems

Though Gentry's blueprint[1] provides a solution, what remains is developing the basic scheme to have more feasible ones. The major drawback of the schemes based on Gentry's blueprint has been large public key size, many keys, growth of ciphertext per computation in a circuit and accumulation of noise. While a major application of FHE is delegation of computation due to lack of resources at the user-end, majority of schemes are computationally intensive making it practically of no use for such users. Another open problem is of reducing key sizes to a manageable level since the procedure requires at least three keys (encryption, re-encryption or evaluation, decryption). The approach should be now to focus on devising application-specific homomorphisms, like a light weight scheme, a fast scheme, a semantically secure scheme, a multiparty scheme and so on. The schemes which are currently in use for applications like e-voting, PIR etc are not fully homomorphic. They are either SHE (Somewhat Homomorphic encryption) or are homomorphic over a limited number of circuits/operations, hence limited to a few number of applications, cannot be extended or generalized for complete category of applications. Mostly all schemes proposed so far are based on public-key cryptography. It has obvious advantage of being based on hardness problems like Large Integer Factorization, Diffie-Hellman problems or Approximate GCD problem. But there are applications which inherently would require symmetric keys, or perhaps no use of a public key at all (viz a user storing his private data on cloud only for personal purposes would need only a secret key). Further there are applications oriented towards involvement of more than one party, such as multiparty computation. Majority of the proposed schemes have not explored this area.

Given the large amount of data and huge costs of encrypting and decrypting them (also the large number of keys to be distributed due to multiple stakeholders) gave way to hybrid clouds and data classification. Hybrid clouds allow combining private enterprise clouds with on-premise data (perceived security is high) to collaborate with public clouds involving third party storage providers (not so secure). Data classification involves different levels of security depending on criticality of data. Moreover, many data centric applications involve multiple users and can benefit only if the encryption process can involve the hierarchy of data classification. A possible solution to be explored is incremental encryption with homomorphic properties.

## 1.5 Problem Statement

*Design an efficient and practically feasible fully homomorphic scheme that uses **symmetric** keys and subsequently design a protocol for its use in **multiuser** data-centric applications.*

## 1.6 Our Contribution

Vaikuntanathan presented a state-of-art survey [2] of FHE and how it can be applied for delegation of computation. He raised an open question "Can Homomorphic encryption be efficient enough to be practical?". Our proposal addresses this open problem. The proposed scheme is based on matrix operations which are computationally "light" and fully homomorphic. It uses symmetric keys of small size thereby making it suitable for many data centric applications. It derives its security from hardness of factorizing a large integer, which is basis of many public key cryptosystems. We extend the approach used in [3]. Besides the primitives for encryption, decryption and evaluation, we have included primitives which are useful to adapt the scheme to specific applications of delegating computation and data access control in multiuser environments like that of cloud computing. It can easily be extended for keyword search in indices of encrypted databases, PIR and electronic voting. We have also included a checklist of properties of a homomorphic scheme when employing it to certain application and compared our proposal against it. The proposed scheme is multi-hop, ensures circuit-privacy, can handle arbitrary size of computations without the need of noise management and has scope of parallelization. We also present a formal security analysis of the scheme.

## 1.7 Organization of Dissertation

*Chapter 2* presents a survey of existing homomorphic schemes, the analyses and applications available in literature. It gives a brief and clear overview of the underlying principles of such schemes and their limitations.

*Chapter 3*, as its title – "Homomorphic Encryption" suggests, acquaints with the relevant terminology. It contains formal definitions, properties and representative applications of Homomorphic Encryption.

*Chapter 4*, titled "Fully Homomorphic Encryption Scheme with Symmetric Keys" describes our proposal. it has a detailed description of the primitives, the algorithms along with examples, performance and security analyses.

*Chapter 5* concludes the dissertation and presents the possible aspects in which the further work can be done.

<div style="text-align: right;">**Chapter 2**</div>

# LITERATURE SURVEY

Beginning from the notion of privacy homomorphisms in 1978, Homomorphic Encryption had been more like a holy grail lurking in minds of cryptographers and quest did not end until 2009. Since then the field has been growing rapidly with so much potential anticipated that cryptographers worldwide are now thinking to consider it an entire new field of computer science.

## 2.1 Privacy Homomorphisms

The idea of using homomorphism along with encryption was introduced by Rivest, Adleman, and Dertouzous in 1978 [4]. They asked for an encryption function that permits encrypted data to be operated on, without preliminary decryption of the operands, and they called those schemes privacy homomorphisms. Unfortunately, shortly after its publication, major security flaws were found in the original proposed schemes of Rivest et al. The search for fully homomorphic cryptosystems began.

## 2.2 RSA- A Multiplicative Homomorphic Scheme

In 1978, Rivest, Shamir, and Adleman published their public-key cryptosystem [5], which only uses elementary ideas from number theory. It is one of the first homomorphic cryptosystems. It is the most widely used public-key cryptosystem. It may be used to provide both secrecy and digital signatures and its security is based on the intractability of the integer factorization problem. The RSA scheme has a multiplicative homomorphic property. This means it is possible to perform multiplications with the encryptions of messages without losing or tampering with their underlying information. This is possible since the operation "multiplication" in the ciphertext space ($Z_n$, $\cdot$) can be compared with the operation "multiplication" in the plaintext space ($Z_n$, $\cdot$). The same is illustrated in Fig 2.1.

$$\begin{aligned} c_1 &= m_1^e \bmod n \\ c_2 &= m_2^e \bmod n \\ \hline c_1 \cdot c_2 &= m_1^e \cdot m_2^e \bmod n \quad = (m_1 \cdot m_2)^e \bmod n \end{aligned}$$

**Figure 2.1 Multiplicative Homomorphic Property of RSA cryptosystem**

## 2.3 Paillier – An Additive Homomorphic Scheme

Pascal Paillier introduced his cryptosystem[6] in 1999. The proposed technique is based on composite residuosity classes, whose computation is believed to be computationally difficult. It is a probabilistic asymmetric algorithm for public key cryptography and inherits additive homomorphic properties, specifically the product of two ciphertexts will decrypt to the sum of their plaintexts. It is illustrated in figure.

$$\begin{array}{rcl}
c_1 & = & g^{m_1} x_1^n \bmod n^2 \\
c_2 & = & g^{m_2} x_2^n \bmod n^2 \\
\hline
c_1 \cdot c_2 & = & g^{m_1} x_1^n \cdot g^{m_2} x_2^n \bmod n^2 \quad = g^{m_1+m_2}(x_1 x_2)^n \bmod n^2
\end{array}$$

**Figure 2.2 Additive Homomorphic Property of Pailier Cryptosystem**

## 2.4 Gentry – An Algebraically Homomorphic Scheme

In a breakthrough work Gentry described in 2009 the first encryption scheme that supports both addition and multiplication on ciphertexts, i.e. a fully homomorphic encryption scheme [1]. Gentry used a method which no other researcher tried before. Instead of directly creating a superior scheme, he build one from a somewhat homomorphic scheme, if its decryption circuit is sufficiently simple. The construction proceeds in successive steps: first Gentry describes a "somewhat homomorphic" scheme that supports a limited number of additions and multiplications on ciphertexts. This is because every ciphertext has a noise component and any homomorphic operation applied to ciphertexts increases the noise in the resulting ciphertext. Once this noise reaches a certain threshold the resulting ciphertext does not decrypt correctly anymore; this limits the degree of the polynomial that can be applied to ciphertexts. Secondly Gentry shows how to "squash" the decryption procedure so that it can be expressed as a low degree polynomial in the bits of the ciphertext and the secret key (equivalently a circuit of small depth). Then the breakthrough idea consists in evaluating this decryption polynomial not on the bits of the ciphertext and the secret key (as in regular decryption), but homomorphically on the encryption of those bits. Then instead of recovering the bit plaintext, one gets an encryption of this bit plaintext, i.e. yet another ciphertext for the same plaintext. Now if the degree of the decryption polynomial is small enough, the resulting noise in this new ciphertext can be smaller than in the original ciphertext; this is called the "ciphertext refresh" procedure. Given two refreshed ciphertexts one can apply again the homomorphic operation (either addition or multiplication), which was not necessarily possible on the original ciphertexts because of the noise threshold. Using this "ciphertext refresh" procedure the number of permissible homomorphic operations becomes unlimited and we get a fully homomorphic encryption scheme. The prerequisite for the "ciphertext refresh" procedure is that the degree of the polynomial that can be evaluated on ciphertexts exceeds the degree of the decryption polynomial (times two, since one must allow for a subsequent addition or multiplication of refreshed ciphertexts); this is called the "bootstrappability" condition. Once the scheme becomes bootstrappable it can be converted into a fully homomorphic encryption scheme by providing the encryption of the secret key bits inside the public key.

## 2.5 Improvements to Gentry's Blueprint – Lattice based

### 2.5.1 Implementation of Gentry's blueprint, 2010

At PKC 2010 Smart and Vercauteren [7] made the first attempt to implement Gentry's scheme using a variant based on principal ideal lattices and requiring that the determinant of the lattice be a prime number. However the authors of [7] could not obtain a bootstrappable scheme because that would have required a lattice dimension of at least $n = 227$, whereas due to the prime determinant requirement they could not generate keys for dimensions $n > 2048$, which is essential for security purposes. This implied that Gentry's blueprint was not yet practical.

### 2.5.2 Gentry-Halevi Scheme 2010

The authors in [8] follow the same direction as Smart and Vercauteren[7] , but for key generation they eliminate the requirement that the determinant is a prime. Additionally they present many clever optimizations. Four concrete parameter settings are provided, from a "toy" setting in dimension 512, to "small", "medium" and "large" settings of dimensions 2048, 8192 and 32768, respectively. For the "large" setting public key size is 2.3 Gigabytes. The authors of [8] report that for an optimized implementation on a high-end workstation, key generation takes 2.2 hours, encryption takes 3 minutes, and ciphertext refresh takes 30 minutes.

### 2.5.3 Optimized Gentry, 2010

Concurrently, Stehle and Steinfeld described two improvements [9] to Gentry's fully homomorphic scheme based on ideal lattices and its analysis. They introduced a probabilistic decryption algorithm that can be implemented with an algebraic circuit of low multiplicative degree. Combined together, these improvements lead to a faster fully homomorphic scheme, with a $\tilde{O}(\lambda^{3.5})$ bit complexity per elementary binary add/mult gate.

### 2.5.4 SIMD Gentry, 2011

Gentry's scheme [1] performs encryption and decryption on plaintext of 1-bit length. Hence, it is intuitive to think that certain operations could be performed on several bits in parallel to reduce runtime. In [7], Smart and Vercauteren mentioned that SIMD(single instruction, multiple data) style operations on data can be supported by their scheme. In [10], Smart and Vercauteren show how to select parameters for Gentry and Halevi's implementation [8] that use SIMD operations for the somewhat homomorphic scheme, how to construct a fully homomorphic scheme when performing re-encryptions in parallel and in which way SIMD operations can be useful in practice. The main point is that the parallel version is 2.4 times faster than the standard FHE scheme and the ciphertext size is reduced by a factor 1/72. Thus, exploiting parallelism in the constituent algorithms can increase efficiency of a scheme.

### 2.5.5 Gentry-Halevi without squashing, 2011

Gentry and Halevi in [11] show how to get rid of squashing as well, using a completely different technique, while the construction still relies on ideal lattices. The new approach constructs FHE as a hybrid of a SWHE and a multiplicatively homomorphic encryption (MHE) scheme. The new approach shows how to bootstrap without having to "squash" the decryption circuit. The main technique is to express the decryption function of SWHE schemes as a depth-3 (ΣΠΣ) arithmetic circuit of a particular form. When evaluating this circuit homomorphically (as needed for bootstrapping), it uses a MHE scheme, such as Elgamal, to handle the Π part. Due to the special form of the circuit, the switch to the MHE scheme can be done without having to evaluate anything homomorphically. The result is translated back to the SWHE scheme by homomorphically evaluating the decryption function of the MHE scheme. Thus, the SWHE scheme only needs to be capable of evaluating the MHE scheme's decryption function, not its own decryption function, thereby avoiding the circularity that necessitated squashing in the original blueprint.

### *2.5.6 Gentry-Halevi-Smart, 2011*

The main bottleneck in the bootstrapping procedure of Gentry's blueprint is the need to evaluate homomorphically the reduction of one integer modulo another. This is typically done by emulating a binary modular reduction circuit, using bit operations on binary representation of integers. Gentry, Halevi and Smart present a simpler approach[12] that bypasses the homomorphic modular-reduction bottleneck to some extent, by working with a modulus very close to a power of two. The method is easier to describe and implement than the generic binary circuit approach, and is likely to be faster in practice. In some cases it also allows storing the encryption of the secret key as a single ciphertext, thus reducing the size of the public key. This method can also be combined with the SIMD homomorphic computation techniques.

## 2.6 Fully Homomorphic Encryption based on Approximate GCD

### *2.6.1 DGHV, 2010*

Based on Gentry's approach a different fully homomorphic scheme by van Dijk, Gentry, Halevi and Vaikuntanathan (DGHV) over the integers appeared at Eurocrypt 2010 [13]. As in Gentry's scheme the authors first describe a somewhat homomorphic scheme supporting a limited number of additions and multiplications over encrypted bits. Then they apply Gentry's "squash decryption" technique to get a bootstrappable scheme and then Gentry's "ciphertext refresh" procedure to get a fully homomorphic scheme. The main appeal of the scheme (compared to the original Gentry's scheme) is its conceptual simplicity: all operations are done over the integers instead of ideal lattices. However the public-key was in $\tilde{O}(\lambda^{10})$ which is too large for any practical system. The major achievement of [13] over [1] was that now the plaintext consisted of integers rather than single bits leading to a better blueprint to improve upon.

### *2.6.2 Improved DGHV, 2011*

Coron et al [14] showed how to reduce the public key size of the somewhat homomorphic scheme from $\tilde{O}(\lambda^{10})$ down to $\tilde{O}(\lambda^{7})$. The idea consists in storing only a smaller subset of the public key and then generating the full public key on the y by combining the elements in the small subset multiplicatively. The new scheme is still semantically secure, but under a stronger variant of the approximate GCD assumption. The second contribution of [14] is to describe an implementation of the fully homomorphic DGHV scheme under new variant, using some of the optimizations from [8]. They use the refined analysis from [9] of the sparse subset sum problem; but not the probabilistic decryption circuit from [9] because as in [8] the error probability is too high for chosen set of parameters. The main difficulty is to determine a secure set of concrete parameters. The approach in [14] is to implement the known attacks, measure their running time and extrapolate for large parameters, so that concrete parameters according to the desired level of security can be fixed. [14] have obtained similar performances as the Gentry-Halevi implementation [8]. More precisely the four security levels inspired by the levels from [8] (though they may not be directly comparable due to different notions of "security bits"): "toy", "small", "medium" and "large", corresponding to 42, 52, 62 and 72 bits of security respectively. For "large" parameters, encryption and recryption take 3 minutes and 14 minutes respectively, with a public key size of 800 MBytes. Decryption is always close to instantaneous. This shows that fully homomorphic encryption can be implemented with a simple scheme.

### *2.6.3 Attack on DGHV, 2012*

Chunsheng proposed a heuristic attack [15] on the fully homomorphic encryption over the integers by using lattice reduction algorithm. Their result shows that one can directly obtain the plaintext from a ciphertext and the public key without using the secret key for some parameter settings of the FHE in [13]. They constructed a new lattice based on the public key and recover the plaintext bit from ciphertext by applying LLL reduction algorithm. They further showed that such an attack can be avoided by setting parameter $\gamma=\lambda^{6}$. But, the scheme is less practical in this case. In addition, they suggested an improvement scheme to avoid the above lattice attack. The secret key which is a large integer in [13] is now replaced by a matrix. To implement FHE, one only needs to add ciphertexts of the secret key to the public key. The size of the public key is $O(\lambda^3 \log \lambda)$ and the size of the secret key is $O(\lambda^2)$.

### *2.6.4 Batch FHE, 2012*

Coron, Lepoint and Tibouchi [16] extended the DGHV scheme to batch fully homomorphic encryption, i.e. to a scheme that supports encrypting and homomorphically processing a vector of plaintext bits as a single ciphertext. It maintains semantic security and allows one to perform arbitrary permutations on the underlying plaintext vector given the ciphertext and the public key. Though there is no notable achievement in terms of efficiency, it presents a new approach for obtaining features of LWE-based FHE scheme in a scheme based on Approximate-GCD.

### *2.6.5 CRT-based FHE, 2012*

Kim, Lee, Yun and Cheon [17] combined the ideas of [4] and [13]. As compared to [13] this scheme has larger plaintext, reduced computation overhead and support for SIMD style operations. Though the achievement is not on efficiency, [17] suggests new methods to construct a fully homomorphic encryption scheme.

## 2.7 Fully Homomorphic Encryption based on Ring-Learning with Errors

### *2.7.1 FHE-LWE, 2011*

Schemes based on Gentry's blueprint suffer from large size of keys and high per-gate evaluation time which is a bottleneck in practical deployment of FHE. This led to a new series of research works. In particular, Brakerski and Vaikuntanathan [18] show that (leveled) FHE can be based on the hardness of the much more standard "learning with errors" (LWE) problem, using a new re-linearization technique. In contrast, all previous schemes relied on complexity assumptions related to ideals in various rings. Instead of using squashing, this proposal introduced a new dimension-modulus reduction technique, which shortens the ciphertexts and reduces the decryption complexity, without introducing additional assumptions. This scheme has very short ciphertexts and can be used it to construct an asymptotically efficient LWE-based single-server private information retrieval (PIR) protocol, also proposed in [18].

### *2.7.2 BGV, 2011*

Brakerski, Gentry and Vaikuntanathan [19] build on (a refinement of) the main technique in [18] to construct an FHE scheme with asymptotically linear efficiency, that is the per gate computation is almost linear in security parameter. While [18] uses modulus switching in "one shot" to obtain a small ciphertext, this scheme uses modulus switching iteratively, to keep the noise level essentially constant, while sacrificing modulus size and gradually sacrificing the remaining homomorphic capacity of the scheme.

### *2.7.3 Optimized BGV, 2011*

Lauter, Naehrig and Vaikuntanathan published results [20] of implementation of the scheme BV [18]. They also proposed a number of application-specific optimizations to the scheme. Most notably they show how to convert between different message encodings in a ciphertext. They propose two methods. The first is to encode integers in a ciphertext so as to enable efficient computation of their sums and products over the integers. This is useful in computing the mean, the standard deviation and other private statistics efficiently. The second trick shows how to "pack' n encryptions of bits into a single encryption of the n-bit string. Some homomorphic operations, for example comparison of integers or private information retrieval, require bit-wise encryptions of the input. Once the answers are computed, though, they can be packed into a single encryption using this trick.

### *2.7.4 HELib, 2013*

Recently, IBM has released software package HELib in April 2013. HElib is a software library that implements homomorphic encryption (HE). Currently available is an implementation of the BGV [19] scheme, along with many optimizations to make homomorphic evaluation run faster, focusing mostly on effective use of the Smart-Vercauteren [7] ciphertext packing techniques and the Gentry-Halevi-Smart [12] optimizations.

## 2.8 Fully Homomorphic Encryption based on Large Integer Factorization

### *2.8.1 Xiao et al, 2012*

In 2012 Xiao et al developed a novel symmetric-key homomorphic encryption scheme [3]. It is proven that the security of this encryption scheme is equivalent to the large integer factorization problem, and it can withstand an attack with up to $m \ln \text{poly}(\lambda)$ chosen plaintexts for any predetermined $m$, constant that is polynomial in the security parameter $\lambda$. Multiplication, encryption, and decryption are almost linear in $m\lambda$, and addition is linear in $m\lambda$. The scheme downgrades the security requirement to achieve efficiency. Although the algorithm is not semantically secure, it can face an adversary with up to $m \ln \text{poly}(\lambda)$ chosen plaintext and ciphertext pairs, and the security is equivalent to the large integer factorization problem. Thus, homomorphic encryption scheme [3] can be used in applications where semantic security is not required and one-wayness security is sufficient.

A further consideration in [3] is practical multiple-user data-centric applications. To allow multiple users to retrieve data from a server all users need to have the same key. In [3] the master encryption key is transformed into different user keys to develop a protocol to support correct and secure communication between the users and the server using different user keys. The data in the data center are encrypted using homomorphic encryption with a "master key" $k$. Different keys are assigned to different users which are actually transformations of master key $k$. Such multi-user system can withstand an adversary with upto $m \ln \text{poly}(\lambda)$ plaintext-ciphertext pairs.

### *2.8.2 MORE&PORE, 2012*

Kipnis and Hibshoosh [21] present high performance non-deterministic fully-homomorphic methods for practical randomization of data (over commutative ring), and symmetric-key encryption of random mod-N data over ring $Z_N$ well suited for crypto applications. These methods secure, for example, the multivariate input or the coefficients of a polynomial function running in an open untrusted environment. The scheme has matrix based operation very similar to [3] and polynomial based operations which is a novel idea. The efficient nature of the methods - one large-number multiplication per encryption and six for the product of two encrypted values - motivates and enables the use of low cost collaborative security platforms for crypto applications such as keyed-hash or private key derivation algorithms. It is shown how to secure OSS public-key signature against Pollard attack. Further, [21] demonstrates how the homomorphic randomization of data can offer protection for an AES-key against side-channel attacks. Finally, the methods provide both fault detection and verification of computed-data integrity.

## 2.9 Homomorphic Encryption in Cloud Computing

Fully Homomorphic Encryption combines security with usability. It can help preserve customer privacy while outsourcing various kinds of computation to the cloud, besides storage. Some concrete and valuable applications of FHE have been mentioned in [20]. They have considered situations where data streams from multiple sources, is uploaded in encrypted form to the cloud, and processed by the cloud to provide valuable services to the content owner. There are two aspects of the computation considered: the data itself (confidentiality), and the function to be computed on this data (circuit privacy). Depending on whether one or both of these are confidential and hence not to be disclosed to the cloud, [20] proposes three broad kinds of applications:

1. Medical applications: private data, public functions
2. Financial Applications: private data, private functions
3. Advertising and Pricing: Only results are public

Application of FHE to database querying is studied systematically in [22]. It identifies what fully homomorphic encryption can do and what it cannot do well for supporting general database queries at a conceptual level. The study shows that using a fully homomorphic encryption scheme that supports addition, multiplication, AND and XOR on ciphertexts, it is possible to process a complex selection, range, join or aggregation query on encrypted data on the server side, and to return the encrypted matching answers in a result buffer. For queries without fixed answer sizes, it is however not guaranteed all matching answers will be correctly constructed from the result buffer, instead the answers can be constructed from the result buffer with overwhelming probability.

## 2.10 Survey Extraction

Encryption schemes with homomorphic properties would suffice the need of security meanwhile preserving system usability in clouds. While a major application of FHE is delegation of computation due to lack of resources at the user-end, majority of schemes are computationally intensive making it practically of no use for such users. The major problem is of reducing key sizes to a manageable level since the procedure requires at least three keys (encryption, re-encryption or evaluation, decryption). A related issue is the noise management – as the noise associated with the homomorphic evaluation increases with every operation on the ciphertext. Moreover, this noise puts a bound on the size of circuits that can be correctly evaluated homomorphically. Mostly the issue has been resolved by some "refreshing" techniques which in turn increase the overall time complexity of a computation. Much of the research has been devoted towards developing FHE schemes using public-key cryptographic primitives; area of symmetric FHE should also be explored as there are many applications which inherently are suitable for symmetric encryption. Then, there are certain applications which involve multiple users, hence requiring multiparty protocols involving FHE. Majority of the existing FHE schemes have not explored this area.

Furthermore, there is a need of a well-agreed upon list of properties that any homomorphic encryption scheme would possess in order to be deployable practically. The properties like circuit privacy, targeted malleability have been discussed a lot in literature, yet all of them have not been consolidated. Also, security notions for conventional encryption have been as is applied to homomorphic encryption, rendering any homomorphic scheme non-INDCPA2 secure, which could mislead users to think that a homomorphic encryption is not much secure for them, while FHE aims at ensuring security where no other means is able to (like to the data being used in public cloud).

# Chapter 3

# HOMOMORPHIC ENCRYPTION

The aim of homomorphic cryptography is to ensure privacy of data in communication and storage processes, such as the ability to delegate computations to untrusted parties. If a user could take a problem defined in one algebraic system and encode it into a problem in a different algebraic system in a way that decoding back to the original algebraic system is hard, then the user could encode expensive computations and send them to the untrusted party. This untrusted party then performs the corresponding computation in the second algebraic system, returning the result to the user. Upon receiving the result, the user can decode it into a solution in the original algebraic system, while the untrusted party learns nothing of which computation was actually performed. Fig 3.1 illustrates this.

Suppose we have a homomorphic cryptosystem which can translate operations on integers to operations on polynomials of single variable. As shown in Fig 3.1, two integers are encrypted into polynomials $p_1(x)$ and $p_2(x)$. Now when these polynomials are added to give a third polynomial, it is required that the resultant polynomial when translated back should be equal to sum of plaintext integers.

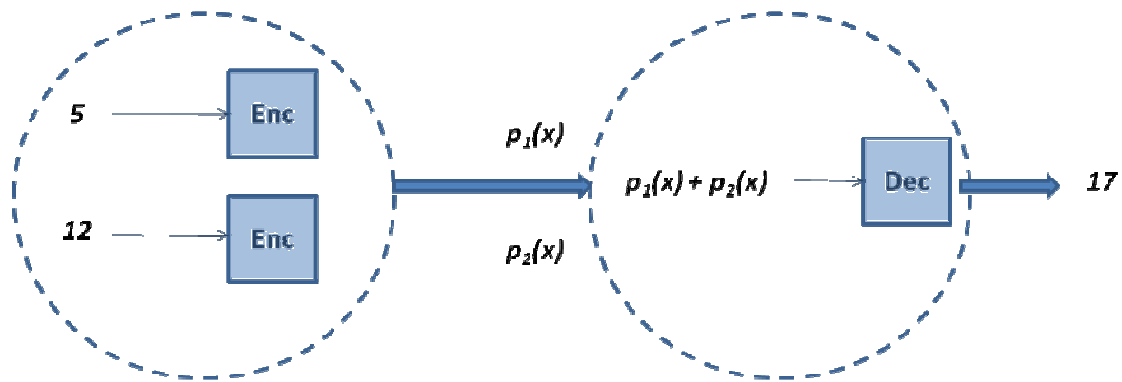

**Figure 3.1 Concept of Homomorphic encryption**

This chapter discusses various basic definitions and other terminology related to homomorphic cryptosystems. Thereafter we present a consolidated list of properties of Homomorphic cryptosystems, and some representative applications.

## 3.1 Homomorphic Encryption: Terminology

At a high-level, the essence of fully homomorphic encryption is simple: given ciphertexts that encrypt the plaintexts $x_1, x_2 \ldots x_n$, fully homomorphic encryption should allow anyone (not just the key-holder) to output a ciphertext that encrypts $f(x_1, x_2 \ldots x_n)$ for any desired function *f*, as long as that function can be efficiently computed. No information about $x_1, x_2 \ldots x_n$ or $f(x_1, x_2 \ldots x_n)$ or any intermediate plaintext values, should leak; the inputs, output and intermediate values are always encrypted. There are different ways of defining what it means for the final ciphertext to "encrypt" $f(x_1, x_2 \ldots x_n)$. The minimal requirement is correctness. Various aspects of such computation along with encryption lead to different forms of Homomorphic encryption.

Below we present formal definitions related to cryptosystems which possess homomorphic computation capability.

*Homomorphic Encryption*

Formally, homomorphic encryption scheme has been defined till now in context of public key systems only. We extend the existing definition so as to incorporate both symmetric as well as public key systems. A homomorphic encryption scheme is a quadruple of probabilistic-polynomial time algorithms

$\varepsilon$=(Keygen, Enc, Dec, Eval)

Keygen – In public-key based systems the key generation algorithm takes input the security parameter $\lambda$ and outputs keys (*pk, sk, ek*), where *pk* is public key *sk* is private key, and *ek* is evaluation key. In symmetric key systems algorithm takes input the security parameters $\lambda$ and *m* and outputs keys (*k, ek*) where *k* is secret key and *ek* is evaluation key.

Enc – The encryption algorithm converts plaintext to ciphertext as $c \leftarrow Enc(\pi, key_1)$. $\pi$ is a plaintext bit or integer, and $key_1$ is *pk* for public crytptosystem and *k* for symmetric scheme.

Dec – The decryption algorithm converts ciphertext to plaintext as $\pi \leftarrow Dec(c, key_2)$. $key_2$ is *sk* for public crytptosystem and *k* for symmetric scheme.

Eval – The homomorphic evaluation algorithm evaluates the result of a computation *f* on ciphertexts $c_1, c_2, \ldots, c_l$ using evaluation key *ek*. $c_f \leftarrow Eval(f, c_1, c_2, \ldots, c_l, ek)$. Use of this key is optional, as in some schemes, like that of our proposal, there is no need of an evaluation key.

Here if *f* is represented as an arithmetic circuit or a Boolean circuit equivalently, the scheme is said to be *circuit-based*. When *f* is defined as mathematical function, the scheme is *non-circuit based*.

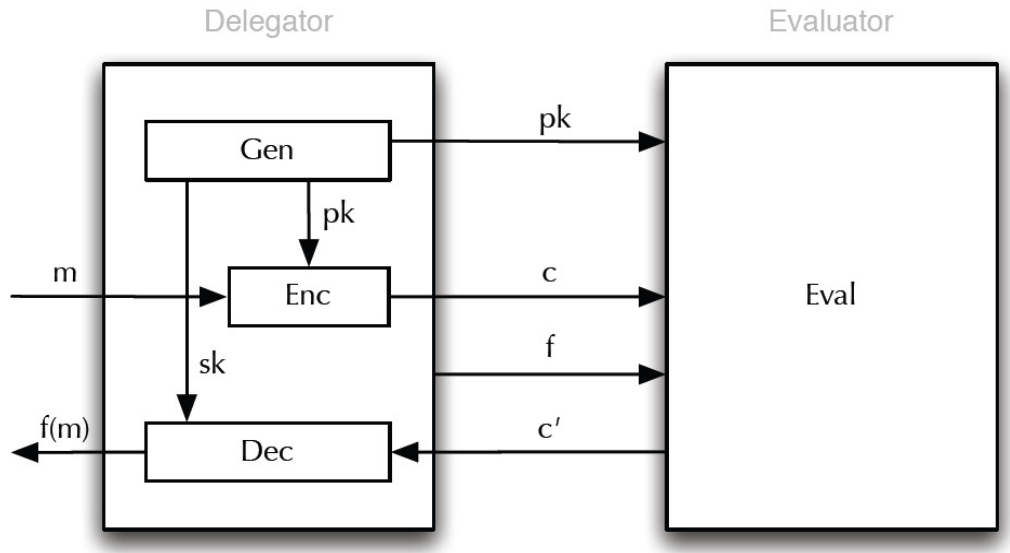

**Figure 3.2 Homomorphic Encryption with Asymmetric keys**

*C-homomorphism*

Let C be a class of functions. A scheme $\varepsilon$ is C-homomorphic if for every function in C, the Eval algorithm of $\varepsilon$ outputs such that $c_f \leftarrow Eval(f, c_1, c_2, ..., c_l, ek)$ and $f(\pi_1, \pi_2, ..., \pi_l) = Dec(c_f, key_2)$

*Compactness*

Compactness requires that the size of the ciphertext after homomorphic evaluation does not depend on either the number of inputs or the complexity of the function *f*, but only on the size of the output of *f*.

*Somewhat homomorphic*

Somewhat homomorphic scheme supports a limited number of additions and multiplications on ciphertexts. This is because every ciphertext has a noise component and any homomorphic operation applied to ciphertexts increases the noise in the resulting ciphertext. Once this noise reaches a certain threshold the resulting ciphertext does not decrypt correctly anymore; this limits the degree of the polynomial that can be applied to ciphertexts.

*Fully homomorphic encryption*

A scheme $\varepsilon$ is fully homomorphic if it is both compact and homomorphic for the class of all arithmetic circuits over GF(2).

*Leveled fully homomorphic*

This is a relaxation in fully homomorphic scheme. A leveled fully homomorphic encryption scheme is a homomorphic scheme where Keygen algorithm gets an additional input $l$ and the resulting scheme is homomorphic for all depth-$l$ binary arithmetic circuits.

*Multi-hop Homomorphic*

In some applications, it is useful to require that the output of algorithm Eval can be used as an input for another homomorphic evaluation. A homomorphic encryption scheme with this property is called a "multi-hop homomorphic" encryption scheme.

## 3.2 Properties of Homomorphic Encryption

Field of homomorphic cryptography is yet in developing stage and various schemes differ in what they have to offer. Selecting a scheme for an application requires a consumer to check its suitability to the application. Various applications require homomorphic operations in different perspective. This variety poses some challenges in designing of fully homomorphic schemes. Also, certain issues should be addressed well before one embarks upon devising a new scheme. This section presents some of the major concerns related to as to what is expected of a fully homomorphic encryption scheme so that it is practical and feasible enough to be employed in certain application. We have identified some common properties which already exist or are desirable in a homomorphic encryption scheme. They are:

1. *Circuit/Function privacy* - Circuit privacy is a property of homomorphic encryption that guarantees that the server's input namely, the computation function $f$ – remains private from the client. In particular, circuit privacy requires that the output of $Eval(f, c_1, c_2, ..., c_l, ek)$ does not reveal any information about $f$ to the client, beyond the output $f(\pi_1, \pi_2, ..., \pi_l)$. This has a security implication when computations have been delegated to an untrusted party. Hence, applications involving multiparty computation, secure delegation of computation, applications involving public data with private functions and the likes.

2. *Targeted Malleability* –In context of conventional cryptography, malleability is an undesirable property. An encryption algorithm is malleable if it is possible for an adversary to transform a ciphertext into another ciphertext which decrypts to a related plaintext. That is, given an encryption of a plaintext m, it is possible to generate another ciphertext which decrypts to f(m), for a known function f, without necessarily knowing or learning m. It actually allows an attacker to modify the contents of a message. But, isn't this the aim of FHE to be able to compute a function of m without knowing m? Hence, here we regress to a related property called targeted malleability. This notion of "targeted malleability" was put forward by Boneh et al [23]: "given an encryption scheme that supports homomorphic operations with respect to some set of functions F, we would like to ensure that the malleability of the scheme is targeted only at the set F. That is, it should not be possible to apply any homomorphic operation other than the ones in F." [23] also suggests how this can be achieved by requiring the entity performing the homomorphic operation to embed a proof in the ciphertext showing that the ciphertext was computed using an allowable function. The decryptor can then verify the proof before decrypting the ciphertext. The problem that might arise due to repeatedly performing a homomorphic operation is that number of proofs grows making the ciphertext grow at least linearly with the number of repeated homomorphic operations. Thus, limiting this expansion is also highly desirable.

3. *Verifiability of Computation* - Homomorphic encryption is being considered as an answer to the problem of securely outsourcing computations, yet it is useful only when the returned result can be trusted. Here verifiability of the result comes into picture. One such scenario is of delegating computation to a cloud. One of the main security issues that arises in this setting is how can the clients trust that the cloud performed the computation correctly? After all, the cloud has the financial benefits to run a fast computation which could be incorrect, freeing up valuable compute time for other transactions. Is there a way to verifiably outsource computations, where the client can, without much computational effort, check the correctness of the results provided by the cloud? Furthermore, can this be done without requiring much interaction between the client and the cloud?

4. *Multiple Users* - Almost every proposed fully homomorphic scheme considers single user setting, with the exception of [3] which discusses how multiple users can participate in a homomorphic operation. Most of the applications of homomorphic operations involve multiple data coming from different users (like multiparty computations). Hence, the focus should be on devising homomorphic operations which can be used in multiuser systems.

5. *Parallel Computations* – If the evaluation function of a homomorphic scheme is able to use the inherent parallelism of certain computations like matrix multiplication, we can benefit in terms of efficiency. If a number of bits can be packed into single argument (like an integer) and the homomorphic evaluation function performs as if the function was performed per bit, we can benefit in terms of communication cost, length of ciphertext and efficiency. This promises to be a field of interesting research. The approach of Single Instruction Multiple Data can be applied when many messages are to be encrypted using same key, or decrypted.

6. *Unlinkability* - A term related to circuit privacy is unlinkability, which asks that the output of the homomorphic evaluation algorithm is computationally indistinguishable from the output of the encryption algorithm.

7. *Multi-hop* – If a sequence of operations can be performed on a ciphertext in succession without the need of any intermediate step which is not a part of overall computation being performed, then the FHE scheme is said to be multi-hop. It is possible if output of the Eval algorithm is of the same form as its input.

## 3.3 Applications of Homomorphic Encryption

Potential of homomorphic encryption had been identified very early. Since then there have been many applications which necessitated a scheme that could compute homomorphically on encrypted data. But with the growing interest and inclination towards cloud computing has opened numerous possible application areas for HE. According to authors in [20] these applications can be majorly classified based on whether we expect confidentiality of data or circuit privacy or both. The categories are:

- Private Data, Public functions: like in Medical Applications
- Private data, Private functions: like in Financial Applications
- Applications like Advertising and pricing where only results should be public

All the above categories of applications assume single data (content) owner who encrypts the data and stores it on an untrusted cloud. But with different cloud models and usage scenarios upcoming we should look at a few representative categories of applications of Homomorphic Encryption.

### 3.3.1 Electronic Voting

Electronic voting is a special case of delegation of computation where one would like the election authorities to be able to count the votes and present the final results, but dislikes the idea that individual votes are first decrypted and afterwards tallied. In a voting system based on homomorphic encryption voters take turns incrementing an encrypted vote tally using a homomorphic operation. They are only allowed to increase the encrypted tally by 1 (indicating a vote for the candidate) or by 0 (indicating a no vote for the candidate). In elections where each voter votes for one of $\ell$ candidates, voters modify the encrypted tallies by adding an $\ell$-bit vector, where exactly one entry is 1 and the rest are all 0's. They are unable to modify the counters in any other way. Thus, homomorphic encryption is a solution to creating a "secret ballot" system online, where neither votes are disclosed to anybody else except the voter, but also issues like vote-buying and coercing can be dealt with.

### 3.3.2 Spam filters

A spam filter implemented in a mail server adds a spam tag to encrypted emails whose content satisfies a certain spam predicate. The filter should be allowed to run the spam predicate, but should not modify the email contents. In this case, the set of allowable functions F would be the set of allowable spam predicates and nothing else. As email passes from one server to the next each server homomorphically computes its spam predicate on the encrypted output of the previous server. Each spam filter in the chain can run its chosen spam predicate and nothing else.

### 3.3.3 Data management and Query processing in Clouds

If all data (personal, health, financial etc) stored in the cloud were encrypted, that would effectively solve issues related to data security. However, a user would be unable to leverage the power of the cloud to carry out computation on data without first decrypting it, or shipping it entirely back to the user for computation. The cloud provider thus has to decrypt the data first (nullifying the issue of privacy and confidentiality), perform the computation then send the result to the user. What if the user could carry out any arbitrary computation on the hosted data without the cloud provider learning about the user's data - computation is done on encrypted data without prior decryption. This is the promise of Fully homomorphic encryption schemes. However, there has not been a systematic study that analyzes the use of fully homomorphic encryption for solving database queries beyond simple aggregations and numeric calculations, such as selection, range and join queries. Wang et al [22] discuss this and show how to use homomorphic encryption for supporting general database queries at a conceptual level, a scheme that supports addition, multiplication, AND and XOR on ciphertexts can also be used to process a complex selection, range, join or aggregation query on encrypted data on the server side, and to return the encrypted matching answers in a result buffer. It is further observed in [22] that for queries without fixed answer sizes, it is however not guaranteed all matching answers will be correctly constructed from the result buffer; instead the answers can be constructed from the result buffer with overwhelming probability.

### 3.3.4 Multiparty Computation

In the setting of multiparty computations one wants different parties to jointly compute some function without revealing their inputs to each other. Secure multiparty computation (MPC) can be defined as the problem of n players to compute an agreed function of their inputs in a secure way, where security means guaranteeing the correctness of the output as well as the privacy of the players' inputs, even when some players cheat. Presently, to conduct such computations, one entity must usually know the inputs from all the participants; however these computations could occur between mutually untrusted parties, or even between competitors, so if nobody can be trusted enough to know all the inputs, privacy will become a primary concern. This privacy can be achieved through homomorphic encryption, and the computation itself can be expressed as a homomorphic circuit or function.

### 3.3.5 Commitment Schemes

Commitment schemes can be thought of like auctions where the auctioneer wants to assure that the offers are not publicly known in the bidding phase while at the same time ensuring that no one is able to repudiate their own offer. Fully homomorphic operations can be used to find the bidder with maximum offer at any time instant without revealing what the offer is. Also, we need some control mechanism so that bidder himself cannot repudiate or change the offer.

### 3.3.6 Private Information Retrieval

A Private Information Retrieval (PIR) protocol allows a database user, or client, to obtain information from a data- base in a manner that prevents the database from knowing which data was retrieved. Particularly, PIR allows a user to retrieve the $i^{th}$ bit of an n-bit database, without revealing the value of index i to the database. A natural and more practical extension of PIR is PBR(Private Block Retrieval) in which, instead of retrieving only a single bit, the user retrieves a $i^{th}$ block with d bits in it. Though the problem is currently solved by querying dynamically, a homomorphic scheme is a better solution. In particular we consider private information retrieval from either a public database or a database with a group of subscribers. Although clients can download the entire database, this takes too long for a large database. Thus PIR that protects only the user is desirable in this scenario. Currently, PIR using HE focuses on encrypting value of "i", and it is desirable to be able to encrypt the query itself such that the cloud can compute it on encrypted data. Given the rate of developments in homomorphic cryptography this seems achievable.

<div style="text-align: right">**Chapter 4**</div>

# FULLY HOMOMORPHIC ENCRYPTION SCHEME WITH SYMMETRIC KEYS

We describe a fully homomorphic encryption scheme with symmetric keys in this chapter. We also present a formal analysis with proofs for performance and security of this scheme. Later sections of the chapter describe a protocol that uses the scheme for private data processing application.

## 4.1 Preliminaries

All computations are performed within the ring $Z_N$, where N is a composite number, product of 2m numbers. Let $\lambda$ denote the security parameter in context of making the scheme CPA secure. In order to make the scheme withstand $\eta$ number of plaintext attacks we choose m and $\lambda$ such that $\eta = m \ln \text{poly}(\lambda)$, where $\text{poly}(\lambda)$ denotes a fixed polynomial in $\lambda$.

We choose 2m odd numbers $p_i$ and $q_i$, $1 \leq i \leq m$, which are mutually prime and of size $\lambda/2$ bits. If $\lambda$ is sufficiently large we can easily choose $p_i$ and $q_i$, that is we take m to be a polynomial in $\lambda$ to ensure enough primes of length $\lambda/2$ bits. Let $f_i = p_i q_i$ and $N = \prod_{i=1}^{m} f_i$.

***Lemma 1:*** *Given m and $\lambda$ where $m = O(\text{poly}(\lambda))$, it is possible to obtain 2m odd mutually prime numbers of length $\lambda/2$ bits in polynomial time.*

**Proof:** By Prime Number Theorem, there are approximately $\frac{x}{\ln x}$ prime numbers $p \leq x$. Consider primes of length b bits, then there are $\frac{2^b}{\ln 2^b}$ primes of length maximum b bits. Thus, total number of primes of length exactly b bits are $\frac{2^b}{\ln 2^b} - \frac{2^{b-1}}{\ln 2^{b-1}} \approx \frac{1}{\ln 2}\left[\frac{2^b}{b} - \frac{2^{b-1}}{b-1}\right] = \frac{1}{\ln 2} \cdot \frac{b-2}{b(b-1)} \cdot 2^{b-1}$.

Thus, when we are finding 2m primes of length b bits, at any point there are at least $\frac{1}{\ln 2} \cdot \frac{b-2}{b(b-1)} \cdot 2^{b-1} - 2^m$ primes left of length b bits. Since total numbers of exact length b bits is $2^b - 2^{b-1} = 2^{b-1}$, the probability that a random number chosen is prime is

$$\frac{\frac{1}{\ln 2} \cdot \frac{b-2}{b(b-1)} \cdot 2^{b-1} - 2^m}{2^{b-1}} = \frac{1}{\ln 2} \cdot \frac{b-2}{b(b-1)} - \frac{2^m}{2^{b-1}}.$$

For b= $\lambda/2$, this gives $\frac{1}{\ln 2} \cdot \frac{2\lambda - 8}{\lambda(\lambda-1)} - \frac{m}{\sqrt{2}^{\lambda-4}}$.

If m is a polynomial in $\lambda$ and $m \ll \sqrt{2}^{\lambda-4}$ then this probability is non-negligible. Primality of a number can be checked in polynomial time, and we are choosing 2m numbers that are mutually prime, the test can be done in time $O(m)$ which is actually $O(\text{poly}(\lambda))$.

Further we lay the foundation of the security of the scheme which is derived as a reduction to the hardness of large integer factorization problem.

***Lemma 2:*** *Factoring N in polynomial time is infeasible if Large Integer Factorization is infeasible.*

**Proof:** Suppose A is an adversary which can factorize a number n into its two prime factors p and q of approximate equal bit length in polynomial time with probability $p'$. Each factor $f_i$ of N is a number with at least two prime factors. Thus, the probability $p''_i$ that an adversary can factorize $f_i$ is lesser than $p'$. Since, N has m such factors, the probability with which the adversary can factorize N is $\prod_{i=1}^{m} p''_i \leq (p')^m$. If $p'$ is negligible, the probability of factoring N in polynomial time is also negligible.

**Matrices over $Z_N$**

The proposed scheme uses invertible matrices as keys with elements from the ring $Z_N$. Lemma 3 below demonstrates the condition which must be satisfied for a matrix to be invertible in $Z_N$. All matrix operations involved in this scheme are performed modulo N at element level. Calculating inverse of a matrix involves preparing transpose of adjoints (which too are calculated modulo N) and then scalar multiplying by multiplicative inverse in $Z_N$ of the determinant of the matrix.

**Lemma 3:** *A matrix $K \in M_4(Z_N)$ is invertible if and only if $\Delta_K \neq 0$ and $gcd(\Delta_K, N)=1$, where $\Delta_K$ is determinant of K.*

**Proof:** The process of finding inverse of a matrix involves multiplicative inverse of the determinant of the matrix, which exists iff the determinant value and N do not have a factor in common. Also, $\Delta_K \neq 0$ eliminates the division by zero condition.

**Corollary:** No two rows or columns of the matrix K should be linearly dependent for it to be invertible.

This is to ensure that $\Delta_K \neq 0$.

**Lemma 4:** *Let $N = \prod_{i=1}^{m} p_i q_i$, where $p_i$ and $q_i$ are mutually prime odd numbers of length $\lambda/2$ bits, a random matrix $K \in M_4(Z_N)$ is invertible with non-negligible probability.*

**Proof:** As seen in Lemma 3, we need to choose a matrix $K \in M_4(Z_N)$ such that. Since the distribution of values of $\Delta$ of all matrices in $M_4(Z_N)$ is not known, we cannot have a formal proof for this Lemma. Hence we provide an analytical proof. We pick random 100 matrices from the field and check for its invertibility. The experiment is repeated five times for a particular value of N, and is performed for various values of N. Fig 4.1 shows the number of invertible matrices varying with value of N. Fig 4.1(a) shows that for small values of N the probability of a random matrix being invertible is not high, but as we increase N, this probability shoots up. The 2m factors of N here are all prime numbers. Fig 4.1(b) shows when all 2m factors are mutually prime numbers, the probability is as high as is Fig 4.1(a). Hence we can deduce that probability of finding an invertible matrix is non-negligible for large values of N.

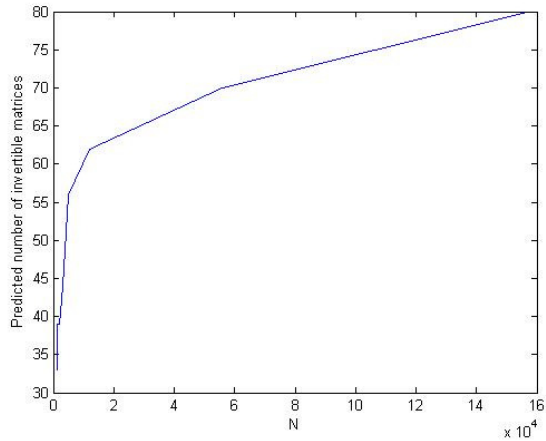

**Figure 4.1(a) Variation of Number of invertible matrices with value of N, all factors prime**

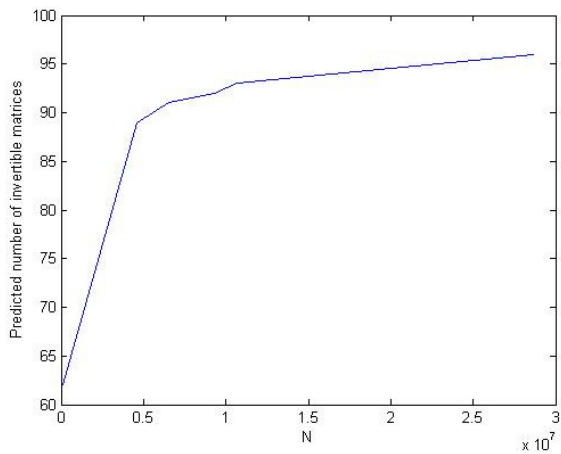

**Figure 4.1(b) Variation of Number of invertible matrices with value of N, all factors mutually prime**

## 4.2 Design Concept

The basic concept is to translate operations on integers in a ring $Z_N$ to operations in ring $M_4(Z_N)$. Thus, all operations are on square matrices of size 4, which are sufficiently small to be used practically. In the context of making a homomorphic scheme to be useful enough, we propose a scheme with following set of operations:

- Cryptographic functions: Functions to generate the symmetric key, encrypting a plaintext, and decrypting a Ciphertext.
- Evaluation operations: To perform any arbitrary operation on data homomorphically, we need to translate it into these basic evaluation operations and then evaluate.

- Application specific functions: These functions provide facilities, like key translate function, recryption etc, for the scheme's adaptation to an application scenario.

The main idea is to construct a matrix with eigenvalue equal to the plaintext *x*. This can be very simply achieved with matrices of size 2, in ring $M_2(Z_N)$ where N=pq, p and q being two large prime numbers, as follows:

$$E_1(x) = \begin{pmatrix} x & 0 \\ 0 & r \end{pmatrix} \mod N$$

The major fallacy here is that x is eigenvalue of eigenvector $\vec{v}_{1,0} = \begin{pmatrix} 1 \\ 0 \end{pmatrix}$. An adversary with a ciphertext has to simply solve a linear equation system $E_1(x).\vec{v}_{1,0} = x.\vec{v}_{1,0}$ to obtain x. to mitigate this problem we can apply a similarity transform to $\begin{pmatrix} x & 0 \\ 0 & r \end{pmatrix}$, governed by an invertible matrix k, called the key. The scheme is now

$$E_2(x,k) = k^{-1} \begin{pmatrix} x & 0 \\ 0 & r \end{pmatrix} k \mod N$$

Though an adversary now cannot establish a linear equation system for transformed eigenvectors, it can very well derive the characteristic equation

$$\det(zI - E_2(x,k)) \equiv 0 \mod N$$
$$\Rightarrow z^2 - (x+r)z + xr \equiv 0 \mod N$$

Though it is infeasible to solve this equation without factorizing N, which is hard. Yet, a chosen plaintext attack is possible by merely two chosen plaintext-ciphertext pairs.

To thwart the chosen plaintext attack, we need to associate x with two eigenvectors $\vec{v}_1$ and $\vec{v}_2$. All plaintexts should have same $v_1$ so that homomorphic operations are possible. Different $\vec{v}_2$ for different plaintext is sufficient to withstand the chosen plaintext attack. We discuss this security aspect formally in a later section. Having two eigenvectors implies increase in the dimension of matrices. Now, we consider matrices of size 4, and also strengthen the scheme by increasing the number of factors of N.

## 4.3 The Cryptosystem

### *4.3.1 Key generation*

The secret key of the proposed scheme is an invertible matrix in $Z_N$, of size 4 hence does not involve any computation theoretically. Yet the process of choosing the key matrix requires more elaboration. We can easily use Lemma 3 and its corollary to check invertibility of a matrix. There can be three approaches:

1. Sequentially search the entire space of all possible 4x4 matrices in $Z_N$ beginning at any random point and stop at the first invertible matrix, returning this as key. The random point of starting the search could be very crucial in deciding the time taken for search.
2. Randomly pick a matrix, check if it is invertible then search is over. Or repeat until an invertible matrix is found. The complexity is that of calculating determinant of a matrix.
3. Instead of just randomly picking a matrix and checking whether it is invertible or not, or going through an entire list, we have a middle path. We build the matrix by random elements and pause as soon as its non-invertibility is proved, restart all over. This has lesser computational cost than finding determinant of entire matrix since it uses the results of lemma 3 and its corollary.

In effect the key matrix in our scheme is constructed using elements from a pseudo-random sequence of numbers, convert them to modulo N. Matrix is constructed row wise, discarding any element (not the entire matrix) which would make a row linearly dependent on a previous row. Also, a row or column of all zeros is avoided. During construction of last row we also check for columns to be linearly independent from each other. When an invertible matrix is found it is published as key. Even this approach matches time complexity of 2[nd] approach in worst cases. This approach is used to implement the 3[rd] step in KeyGen algorithm, as shown below:

> **Keygen$_4$(m, $\lambda$)**
>
> 1. Choose 2m odd numbers $p_i$ and $q_i$, $1 \le i \le m$, which are mutually prime and of size $\lambda/2$ bits.
> 2. Let $f_i = p_i q_i$ and $N = \prod_{i=1}^{m} f_i$.
> 3. Pick an invertible matrix k of size 4, $k \in M_4(Z_N)$
> 4. Compute its inverse as k$^{-1}$ modulo $Z_N$.
> 5. Output $\langle f_i, N, k, k^{-1} \rangle$ as Ktuple.

*4.3.2 Encryption/Randomization*

The plaintext $x \in Z_N$ is encrypted or randomized into a matrix $C \in M_4(Z_N)$. It is indeed a similarity transformation of a diagonal matrix $(x, x_1, x_2, x_3)$, where $x_i, 1 \le i \le 3$ are solutions to sets of linear congruences computed using Chinese Remainder Theorem. The congruences depend on plaintext $x$ and a random values $r$, $r \in Z_N, r \ne x$. First we construct a 3x3 matrix as

$$\begin{bmatrix} x & r & r \\ r & x & r \\ r & r & x \end{bmatrix}$$

Now we pick m rows at random from this matrix to construct a mx3 matrix and call it X. This ensures that each row of X has only one element equal to plaintext x. each column of X is used to form the simultaneous congruences.

The algorithm is as follows:

> **Enc$_4$(x,Ktuple)**
>
> 1. Choose a random value $r \in Z_N, r \ne x$
> 2. Construct a matrix $X(m \times 3)$ such that each row has only one element equal to x, and other two equal to r.
> 3. Using Chinese Remainder Theorem set $x_j, 1 \le j \le 3$ to be solution to the simultaneous congruences $x_j \equiv X_{ij} \mod f_i, 1 \le i \le m$.
> 4. Ciphertext $C = k^{-1} * diag(x, x_1, x_2, x_3) * k$

*4.3.3 Decryption*

This is a single step process which involves applying inverse transformation on the Ciphertext matrix and then extracting the plaintext as first element of the diagonal matrix obtained.

> **Dec**(C, Ktuple$_{2-4}$)
>
> Output the plaintext as $x = (kCk^{-1})_{11}$.

The correctness of the decryption algorithm is proven below.

***Lemma 5: Encryption scheme (KeyGen, Enc, Dec) is correct.***

**Proof:** We know that $k * k^{-1} = k^{-1} * k = I$ and for any matrix A, A*I=I*A=A.

It is easy to note that Dec(C, k)  $= (k * C * k^{-1})_{11}$

$$= (k * k^{-1} * diag(x, x_1, x_2, x_3) * k * k^{-1})_{11}$$

$$= (diag(x, x_1, x_2, x_3))_{11}$$

$$= x$$

This proves the lemma.

**Example 4.1**

We present here a toy example for illustration of the method by selecting lowest prime numbers. Let m=2, let p={3,5} and q={7,11}. This gives $f_1$= 21, $f_2$ = 55 and N= 1155.

For these parameters, suppose the Keygen function generates a key

$$k = \begin{bmatrix} 333 & 1009 & 1093 & 394 \\ 566 & 870 & 285 & 192 \\ 305 & 642 & 456 & 407 \\ 326 & 1103 & 363 & 837 \end{bmatrix}$$

with inverse as

$$k^{-1} = \begin{bmatrix} 33 & 929 & 342 & 393 \\ 963 & 100 & 1113 & 161 \\ 202 & 88 & 1042 & 976 \\ 906 & 1051 & 944 & 441 \end{bmatrix}$$

To encrypt plaintext x=257, we construct diagonal matrix as follows:

Select random number r = 291. The matrix X is

$$X = \begin{bmatrix} 291 & 257 & 291 \\ 291 & 291 & 257 \end{bmatrix}$$

This gives us the linear congruences as follows:

$291 \bmod 21 \equiv x_1 \quad 257 \bmod 21 \equiv x_2 \quad 291 \bmod 21 \equiv x_3$
$291 \bmod 55 \equiv x_1, \quad 291 \bmod 55 \equiv x_2, \quad 257 \bmod 55 \equiv x_3$

The solution to congruences are 291, 236 and 312 respectively. Encryption proceeds as:

$$C = k^{-1} * diag(257, 291, 236, 312) * k = \begin{bmatrix} 464 & 206 & 422 & 308 \\ 585 & 467 & 885 & 945 \\ 957 & 752 & 1119 & 882 \\ 315 & 1136 & 270 & 201 \end{bmatrix}$$

Decryption is done as $x = [k * C * k^{-1}]_{11} = \begin{bmatrix} 257 & 0 & 0 & 0 \\ 0 & 291 & 0 & 0 \\ 0 & 0 & 236 & 0 \\ 0 & 0 & 0 & 312 \end{bmatrix}_{11} = 257$

### *4.3.4 Evaluation*

There is only one general evaluation function defined for computation *f*. It is expected that *f* be translated into basic operations on integers. Actual implementation involves analogous operations on matrices. Namely, to perform addition/subtraction/multiplication/division of two numbers homomorphically, we add/subtract/multiply/divide their ciphertexts simply as two matrices.

$Y \leftarrow \text{Eval}(f, C_1, C_2...C_n)$ performs computation *f* on operands $C_1, C_2...C_n$.

Our evaluation function doesn't require any evaluation key. Note that all operations on matrices are also performed within the ring $M_4(Z_N)$. To illustrate homomorphic operation, we present following example.

### *Example 4.2*

Consider addition of two integers. Let m=2, let p={3,5} and q={7,11}. This gives $f_1$= 21, $f_2$ = 55 and N= 1155. For these parameters, suppose the Keygen function generates the key:

$$k = \begin{bmatrix} 366 & 826 & 315 & 660 \\ 224 & 398 & 457 & 165 \\ 1063 & 849 & 492 & 597 \\ 401 & 1083 & 114 & 496 \end{bmatrix}$$

with inverse as

$$k^{-1} = \begin{bmatrix} 1083 & 213 & 1 & 1053 \\ 1093 & 792 & 84 & 342 \\ 307 & 784 & 877 & 471 \\ 300 & 375 & 874 & 613 \end{bmatrix}$$

To add two numbers, viz 5 and 12, we encrypt them using key k, and obtain following ciphertexts.

$$\text{Enc}(5,k) = C_1 = \begin{bmatrix} 286 & 618 & 534 & 180 \\ 954 & 662 & 651 & 765 \\ 122 & 1131 & 428 & 450 \\ 825 & 285 & 1020 & 196 \end{bmatrix}, \quad \text{Enc}(12,k) = C_2 = \begin{bmatrix} 877 & 813 & 768 & 1122 \\ 60 & 1149 & 1152 & 33 \\ 507 & 245 & 304 & 759 \\ 472 & 531 & 828 & 150 \end{bmatrix}$$

Now, we add the ciphertexts instead of adding plaintexts.

$$C_1 + C_2 = \begin{bmatrix} 8 & 276 & 147 & 147 \\ 1014 & 656 & 648 & 798 \\ 629 & 229 & 732 & 54 \\ 142 & 816 & 693 & 346 \end{bmatrix} = C$$

The resultant C is treated as a ciphertext and decrypted as usual.

$$y = (k.(C_1 + C_2)k^{-1})_{11} = \begin{bmatrix} 17 & 0 & 0 & 0 \\ 0 & 408 & 0 & 0 \\ 0 & 0 & 1079 & 0 \\ 0 & 0 & 0 & 238 \end{bmatrix}_{11} = 17$$

As can be easily observed that the decryption (17) is actually the result of addition of the two plaintexts (5 and 12). Thus illustrating that the scheme is additively homomorphic.

Now, we multiply the ciphertexts C1 and C2.

$$C_1 * C_2 = \begin{bmatrix} 265 & 150 & 180 & 897 \\ 180 & 1005 & 450 & 933 \\ 185 & 775 & 500 & 609 \\ 82 & 816 & 933 & 360 \end{bmatrix} = C$$

The resultant C is treated as a ciphertext and decrypted as usual.

$$y = (k.(C_1 + C_2)k^{-1})_{11} = \begin{bmatrix} 60 & 0 & 0 & 0 \\ 0 & 440 & 0 & 0 \\ 0 & 0 & 673 & 0 \\ 0 & 0 & 0 & 957 \end{bmatrix}_{11} = 60$$

As can be easily observed that the decryption (60) is actually the result of multiplication of the two plaintexts (5 and 12). Thus illustrating that the scheme is multiplicatively homomorphic.

Thus, our scheme is fully homomorphic.

### 4.3.5 Selecting N

We assume a security parameter $\lambda$ in context of making the scheme IND-CPA secure, that is in order to withstand $\eta$ number of plaintext attacks we choose $m$ and $\lambda$ such that $\eta = m \ln \text{poly}(\lambda)$, where $\text{poly}(\lambda)$ denotes a fixed polynomial in $\lambda$. N is computed as product of 2m numbers. Xiao et al in [3] propose all these 2m numbers to be distinct and prime. We modify or relax this requirement to have 2m mutually prime numbers. Moreover, [3] allows N to be even (that is prime number 2 is allowed), which is dropped in our scheme. Thus, N is product of 2m numbers which are odd, mutually prime.

Furthermore, the fact that these 2m numbers could now be composite implies that total number of prime factors of N is more than 2m, thus making scheme more secure (now N needs to be factorized into 2m composite factors). As we will discuss in a later section that security of the scheme is derived from the hardness of the problem of factorizing a large integer. The original scheme [3] has a vulnerability that capturing few ciphertexts may disclose the approximate length of N in bits. In order to benefit from hardness of factorization we increase value of m, implying that we increase the number of factors of N. But m is bound by length of N, hence to increase the number of prime factors of N, we can make these 2m factors composite. Thus, we will have benefit of increasing the number of prime factors of N without increasing m.

## 4.4 Application-specific Primitives

The proposed scheme has primitives which would be useful when using this scheme for delegation of computation. These primitives derive their functionality from properties of matrices but are homomorphic in nature. They can be combined with homomorphic encryption or other encryption techniques for certain practical applications.

### *4.4.1 Lock-Unlock Operations*

In order to evaluate a function homomorphically we need all inputs to be encrypted using same key. Moreover, decryption should also be performed using same key for retrieving result. This leads to natural asking for a method to calculate homomorphically on ciphertexts encrypted by altogether different keys, or atleast related keys but not same key. In this section we introduce primitives which can be used to convert ciphertexts from one key to ciphertext of other key. Also, how to generate a set of related keys so that operations can be performed on ciphertexts encrypted by them can be used in computations in some order and final result is decrypted easily using yet another key.

For any matrix A and an invertible matrix k, Lock operation is defined as

**Lock**(*A,* Ktuple$_{2-4}$)
Output the matrix as $B = k^{-1} * C * k$.

Thus, Lock outputs a randomization of the input matrix under k. It is same as the last step of encryption algorithm. Analogously, Unlock inverts this similarity transformation as in decryption algorithm.

**Unlock**(*B,* Ktuple$_{2-4}$)
Output the matrix as $A = k * C * k^{-1}$.

The exact application is discussed in next chapter. The beauty of this scheme is its simplicity and adaptability to a multikey scenario. Also, the operations are not to be performed necessarily in the order of Lock and then Unlock. We can also have Unlock followed by Lock (ie Unlock(B,k) to obtain A and then Lock(A,k) to get B.)

*4.4.2 Key Set Generation*

For certain multi-user scenarios we need symmetric keys of different levels, in other words we need individual and group keys separately. Yet we desire to have some interoperability among these. Function KeySetGen generates the key matrices which can be used for encryption and Lock-Unlock operations with the property that it produces a set of matching keys.

A set $(k, k', k'', k''')$ is said to be a set of matching keys if $k = k'*k''*k'''$ holds, where k is generally referred to as a master key. This is a three level set. We may also use a two level set in certain applications, that is $(k, k', k'')$ where $k = k'*k''$

---

KeySetGen$_s$($l, m, \lambda$)
1. Choose 2m odd numbers $p_i$ and $q_i$, $1 \leq i \leq m$, which are mutually prime and of size $\lambda/2$ bits.
2. Let $f_i = p_i q_i$ and $N = \prod_{i=1}^{m} f_i$.
3. Pick $l$ invertible matrices $(k^{(1)}, k^{(2)}, ..., k^{(l)})$ of size s, $k^{(i)} \in M_s(Z_N)$
4. Compute $k = \prod_{i=1}^{l} k^{(i)}$
5. Output $\langle f_i, N, k, (k^{(1)}, k^{(2)}, ..., k^{(l)}) \rangle$ as Keyset.

---

This notion of matching keys is also useful when we want to link computations in some order, and at every step of computation the input argument is encrypted using a different key.

## 4.5 A Multiparty Protocol for Privacy Preserving Data Processing

*4.5.1 System Model*

We consider a system model similar to [24] where cooperation of several entities makes the arbitrary times of the homomorphic calculations more efficient. We drop the assumption of non-colluding entities; rather we present a scheme which is collusion-resistant. We assume the processing to be done on data as arithmetic operations within ring $Z_N$.

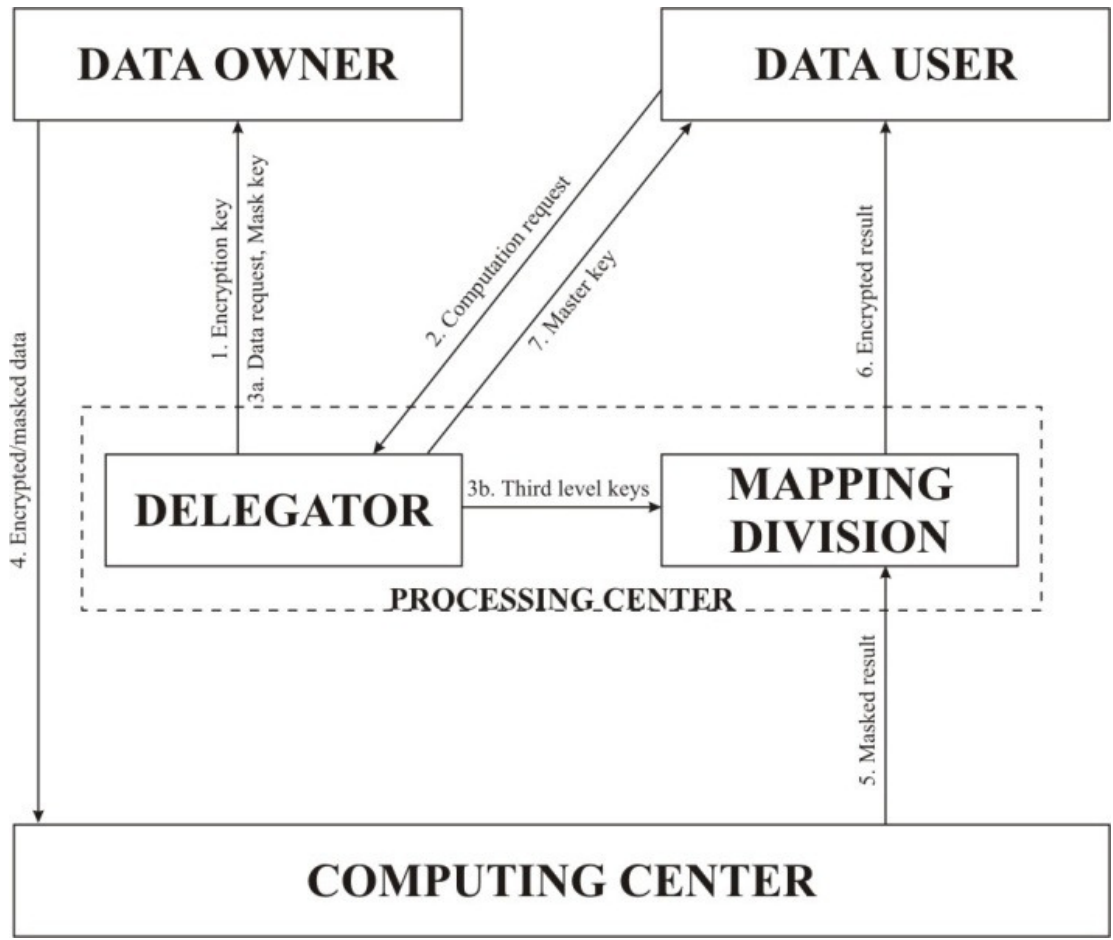

**Figure 4.2 Privacy Preserving Data Processing: System Model and Protocol**

The role and functions of entities (refer Fig 4.2) involved in data processing are as follows:

1. Data owner- possesses raw data which is not disclosed to others. Data owner is responsible for encrypting and masking data and has both encrypt and lock functionality.
2. Processing center- has two divisions. The delegator division tells which data is required for certain computation and how to mask it. It has keyset generation functionality. The mapping division maps results from computation center to be consumed by certain user. It has only Lock functionality and obtains key from Delegator division.
3. Computation center- performs calculations which are requested by the data user. It has access to masked data but not keys. It receives ciphertexts and formula. Sends final result to mapping division of processing center. It has only evaluation functionality.

4. Data user – has intention of performing some computation on data currently owned by data owner. Data user receives only the final result and cannot know any intermediate result or raw data. It has functionality of decryption.

Our goal is computing $f(P)$ where $f$ is a function compose of addition and multiplication, and $P=(x_1,x_2,…x_n)$ is input data. We need to compute the result while keeping P secret. Also, function f is divided into several additions and multiplications, and computation is executed step by step (Note that this is same as Eval function).

Security of the scheme is intuitive as entity which can access encrypted data does not have decryption key, and entity with decryption key are prohibited from accessing encrypted data. In a public key cryptosystem this arrangement would be susceptible to a collusion attack if two entities possessing decryption key and encrypted data decide to collude. But as we will see here these keys are not same in this scheme and hence it is collusion-resistant.

*4.5.2 Multiparty Protocol*

The protocol (refer Fig 4.2) for evaluating a function $f(x_1,x_2,…x_n)$ is:

1. Data owner has data encrypted by key $k^{(1)}$, as $Y_i \leftarrow \text{Enc}(x_i, k^{(1)})$
2. Delegator prepares a list of data which is required for computation and sends it to data owner. Actually it can send the required indices 1…n. [Note that the sent indices have been renumbered for comprehensibility as 1…n and need not be actually continuous.]
3. Data owner masks the data as $Z_i \leftarrow \text{Lock}(Y_i, k^{(2)})$
4. Computation center performs $f$ to produce result as $Z \leftarrow \text{Eval}(f, Z_1, Z_2...Z_n)$ and sends it to Mapping division.
5. Mapping division converts Z as $Y' \leftarrow \text{Lock}(Z, k^{(3)})$
6. Data user retrieves result as $y \leftarrow \text{Dec}(Y', k)$

Here, the keys $k^{(1)}$, $k^{(2)}$, $k^{(3)}$ and k are matching keys with k as master key, generated by KeySetGen.

As can be observed no key is sent to the Computation Center. The keys of data user and mapping division if combined cannot reveal $k^{(1)}$, the actual encryption key. Thus, the protocol is resistant to collusion.

## 4.6 Performance

### *4.6.1 Complexity of Algorithms*

We need to choose $2m$ primes in the encryption scheme. The encryption algorithm requires both two matrix multiplications and also an algorithm to solve the $m$ linear congruences that define the values $a$, $b$, and $c$. It takes time $O(m\lambda)$ to construct the solution to these linear congruences. Multiplication has time complexity $O(m\lambda \log m\lambda \log \log m\lambda)$. So the overall complexity for encryption is $O(m\lambda \log m\lambda \log \log m\lambda)$. The decryption algorithm involves only two matrix multiplications, thus having same time complexity.

Considering the complexity of the multiplication and addition algorithms, observe the size of the integers in the ring $\mathbb{Z}_N$. The value $N$ is the product of $m$ numbers of length $\lambda$ bits, so it is approximately an $m\lambda$ bit number. There exist efficient algorithms for multiplication of $b$ bit integers with complexity $O(b \log b \log \log b)$. For $b = m\lambda$ this becomes $O(m\lambda \log m\lambda \log \log m\lambda)$. Addition is linear and thus has complexity $O(m\lambda)$.

### *4.6.2 Computational Overhead*

Homomorphic evaluation of a function is efficient if it has a low computation overhead. The overhead is defined as the ratio of the time taken for a computation homomorphically over ciphertext to the time taken to compute on plaintext. If a computation consists only of addition, adding two integers homomorphically in our scheme implies adding two matrices. This gives a constant overhead of 16, since we have to add two matrices of size 4, containing 16 numbers. If a computation consists only of multiplication, multiplying two integers homomorphically implies multiplying two matrices, which means 64 additions and 64 multiplications. Since N is a b bit number, cost of multiplying two numbers is $O(b^2)$. Thus giving computation overhead O(b) or $O(m\lambda)$.

Hence, we conclude that our scheme has a worst case computation overhead $O(m\lambda)$ that is varying linearly with the security parameter.

### *4.6.3 Plaintext Expansion*

An integer is encrypted into a matrix of 16 numbers, resulting into a constant expansion factor of 16. It does not vary with bit length of N, and is independent of other security parameters.

Table 4.1 shows comparison between our proposed scheme and other popular FHE schemes with respect to the performance characteristics.

**TABLE 4.1**

**COMPARISON OF PROPOSED SCHEME WITH OTHER FHE SCEHMES**

|  | **DGHV** | **BGV** | **Our Scheme** |
|---|---|---|---|
| **Key Size** | $O(\lambda^{10})$ | Equal to plaintext | $O(m\lambda)$ |
| **Computation Overhead** | $\tilde{\Omega}(\lambda^{3.5})$ | $\tilde{O}(\lambda^2)$ | $O(m\lambda)$ |
| **Plaintext Expansion** | $O(\log \lambda)$ | $O(\lambda^3)$ | $O(1)$ actually 16 |

## 4.7 Security

We shall discuss security of our scheme in terms of key recovery, onewayness, semantic security and indistinguishability. Then we proceed towards proving that the scheme is CPA secure.

### 4.7.1 Security against Key Recovery

In plain words this means that the knowledge of the cipher text must not allow adversaries to retrieve the key. Since for our scheme ciphertext does not reveal anything about key except its length, security against key recovery amounts to security against brute force attack.

Key for our scheme is a $l \times l$ matrix in ring $Z_N$, which leads to $N^{l^2}$ possibilities of a key matrix. The probability that a random generated matrix is a key is $1/N^{l^2}$. Checking whether a random matrix is the key or not, involves two matrix multiplications which implies $\Omega(l^2)$ operations per multiplication. Given N is b bits long, the complexity of brute force attack is $\Omega(l^2 \cdot 2^{bl^2})$. Table 4.2 gives the equivalent security level for different parameter values. It can be easily observed that our scheme is secure against brute-force attack even with smallest parameters.

**TABLE 4.2**

**BRUTE-FORCE SECURITY OF PROPOSED SCHEME**

| Length of N (in bits) | Equivalent security |
|---|---|
| 10 | $2^{172}$ |
| 16 | $2^{268}$ |
| 18 | $2^{300}$ |

*4.7.2 One-way security*

This implies that given a ciphertext an adversary should not be able to retrieve the corresponding plaintext. Since ciphertext is a randomization of the plaintext, and not a direct linear(or polynomial) function of the plaintext, in order to retrieve plaintext from ciphertext an adversary has to invert the similarity transformation and then only can any other linear algebraic methods can be useful to retrieve plaintext.

Let us assume that certain permutation of identity matrix $K_I$ can be used to invert the transformation by following operation: $C' = K_I C K_I^{-1}$, where C is given ciphertext. To obtain plaintext from C' the adversary must be able to factorize N, that is adversary can retrieve plaintext only by solving congruences using Chinese Remainder Theorem, but for that it needs factors of N. Thus, oneway security of our scheme can be reduced to hardness of factorization N. As per Lemma 2, this cannot be done with a nonnegligible probability. Thus, onewayness security reduces to hardness of Large Integer Factorization problem.

Formally we prove this security using following lemmas.

***Lemma 6:*** *For $1<i<N$, there exists a unique element $k_i$ $GL_4(Z_N)$ so that*

$$k_i = \begin{bmatrix} 0 & 1 & 0 & 0 \\ 1 & 0 & 0 & 0 \\ 0 & 0 & 0 & 1 \\ 0 & 0 & 1 & 0 \end{bmatrix} \mod p_i, k_i = I \mod q_i, k_i = I \mod f_j, j \neq i$$

*where I is the identity matrix in $GL_4(Z_N)$. Additionally, $k_i = k_i^{-1}$.*

**Proof:** The first claim follows directly from Chinese Remainder Theorem, as $p_i$ is a factor of N, and $f_j = p_j q_j$ is also a factor of N. Further, we see

$$k_i^2 = \begin{bmatrix} 1 & 0 & 0 & 0 \\ 0 & 1 & 0 & 0 \\ 0 & 0 & 1 & 0 \\ 0 & 0 & 0 & 1 \end{bmatrix} \text{ which implies } k_i = k_i^{-1}.$$

**Lemma7:** Given plaintext x, key k and random element r, there exists y and random element s such that $E(x,k) = E(y, k_i k)$.

**Proof:** Here we note that the diagonal matrix constructed during encryption of plaintext x is like $X = diag(x, a, b, c)$ and it satisfies the congruences $X = diag(x, a_i, b_i, c_i) \bmod f_i$, so

$$k_i . X . k_i^{-1} \bmod p_i = k_i . diag(x, a_i, b_i, c_i) . k_i^{-1} \bmod p_i = diag(a_i, x, b_i, c_i) \bmod p_i$$

Also, $k_i . X . k_i^{-1} \bmod q_i = I . X . I \bmod q_i = X \bmod q_i$ and similarly $k_i . X . k_i^{-1} \bmod f_j = X \bmod f_j, j \neq i$.

Let the diagonal matrix constructed during encryption of plaintext y be $Y = diag(y, a', b', c')$. Then the set of congruences $Y = diag(a_i, x, b_i, c_i) \bmod p_i$, $Y = X \bmod q_i$, and $Y = X \bmod f_j$ has a unique solution by the Chinese Remainder Theorem. This solution also satisfies $Y = k_i . X . k_i^{-1} \Rightarrow k_i^{-1} . Y . k_i = X$.

This implies $E(x, k) = k^{-1} . X . k = k^{-1} k_i^{-1} . Y . k_i k = E(y, k_i k)$, which proves the Lemma.

By Lemma 7 we deduce that an adversary has no polynomial time method to differentiate between the ciphertexts of two given plaintexts *x* and *y* if key is not known. Hence, the onewayness property of our scheme is established.

### *4.7.3 Indistinguishability*

Intuitively, a symmetric encryption scheme is said to exhibit Indistinguishability property if given a ciphertext of one of the two messages selected by challenger, it should be "hard" for the adversary to guess which of two messages corresponds to the ciphertext. The definition involves a simple game where the adversary is tested for the ability to guess which message is encrypted in a given ciphertext. The IND security game is defined as:

1. Attacker produces two messages $m_0$ and $m_1$.

2. The challenger returns the challenge ciphertext $c = Enc(m_b,k)$, b is 0 or 1.

3. Attacker outputs b'.

Attacker or adversary is a winner if it returns b'=b with probability more than 0.5 in polynomial time.

Since the plaintext space is uniform, that is all plaintext have equal bit length, Indistinguishability implies semantic security. Hence, the proposed scheme is semantically secure.

*4.7.4 Security against Known Plaintext and Chosen Plaintext Attack*

Plaintext attack security captures the notion of an adversary who has the ability to eavesdrop on arbitrary messages between a sender and receiver before attempting to decrypt a message. The difference between known-pliantext attack and chosen-plaintext attack is that latter is adaptive one. The notion of security against the Known plaintext attacks is called indistinguishability under Known ciphertext,IND-KPA, defined as:

1. Challenger runs KeyGen

2. (Query Phase I) Attacker is given access to Enc(.,k) oracle.

3. (Challenge Phase) Attacker produces two messages $m_0$ and $m_1$. The challenger returns the challenge ciphertext $c = Enc(m_b,k)$, b is 0 or 1.

4. (Query Phase II) Same as Query Phase I.

5. Attacker outputs b'.

Attacker or adversary is a winner if it returns b'=b with probability more than 0.5. This game can be repeated polynomial number of times. For the adaptive case, the IND-CPA game is the same, except that Attacker generates the next pair of message only after seeing the previous ciphertext. If an encryption scheme is deterministic (the Enc algorithm is deterministic) then there is a unique, consistent encryption for every message. A deterministic encryption scheme cannot be IND-KPA or IND-CPA secure since we can simply ask for the encryption of the two challenge messages during the oracle access step and compare the oracle's response to the challenge ciphertext. Since the proposed scheme is not deterministic, we can claim it to be IND-CPA secure. Next we prove this.

**Lemma 8:** Given a plaintext x, its encryption C with random number r and a key k, any oracle Enc(.,k) will return C with a probability $\frac{1}{N*3^m}$.

**Proof:** The fact to be noted is that the encryption depends on number r which is chosen at random (in step 1 of encryption) by encryption oracle. Since $r \in Z_N$, the probability that same r is chosen is $\frac{1}{N}$, under a uniform probability distribution for selecting random number. Even when same r is chosen, the probability that same m rows of X will be selected (in step 2 of Encryption) to construct linear congruences is $\frac{1}{3^m}$. Thus, probability of producing same ciphertext for a given plaintext and a key is $\frac{1}{N*3^m}$. Hence, the claim is proved.

From Lemma 8 we can observe that even for the smallest possible values of N and m (respectively 1155 and 2), the probability is 0.0000962. Thus, the scheme is IND-KPA secure.

The proposed scheme is CPA secure if the number of chosen plaintext-ciphertext pairs is less than the number of factors used in linear congruences during encryption. In other words, for $m' \leq m$ the scheme is CPA secure for $m'$ plaintext-ciphertext pairs. This is so because more than m pairs chosen adaptively can help adversary to factorize N, hence break the scheme.

## 4.8 Properties

As discussed in Section 3.2, the properties of a Homomorphic scheme decide the category of applications it can be used in. Hence, it is important to discuss the properties of our proposed scheme in the light of its deployment to practical use. They are:

1. Circuit/function privacy – All intermediate and final results of any computation are element of $M_4(Z_N)$. Hence, the vital information like number of parameters, size of circuit, intermediate results or purpose of function cannot be deduced from the result itself.

2. Multiple Users – Our scheme can be deployed for multiple user computations. We have shown in Section 4.4 a possible method to do so.

3. Parallel computation – The cryptographic primitives as such do not have a scope of parallelization. But, complexity of all algorithms are dependent on the matrix operations performed therein. These can be optimized by parallelizing the algorithms for addition and multiplication.

4. Unlinkability – The output of encryption algorithm is a 4x4 matrix, indistinguishable from the output of the evaluation algorithm, hence is the unlinkability property of our scheme.

5. Multi-hop – The output of algorithm Eval is a 4x4 matrix which can again be input to Eval algorithm without any intermediate (extra) operation, thus making possible multiple "hops" of evaluation to be performed in succession.

## 4.8 Implementation Results

We implement our algorithm using Java and evaluate its execution time. The computations were performed on a 3.40 GHz Intel Core i3-2130 processor. Table 4.3 lists the execution time for key generation, encryption and decryption for various lengths of N.

**TABLE 4.3**

**EXECUTION TIME OF KEY GENERATION, ENCRYPTION AND DECRYPTION FOR VARIOUS LENGTHS OF N**

| |N| (in bits) | Key Generation | Encryption | Decryption |
|---|---|---|---|
| 12 | 33 ms | 31 ms | 17 ms |
| 18 | 356 ms | 193 ms | 124 ms |
| 24 | 16.43 s | 42.91 s | 6.45 s |

The data for homomorphic evaluations was gathered from running 10000 additions and 100 multiplications of randomly selected numbers of varying length. Table 4.4 lists the execution time required for homomorphic addition and multiplication.

**TABLE 4.4**

**EVALUATION TIME OF ADDITION AND MULTIPLICATION FOR DIFFERENT LENGTHS OF N**

| |N| (in bits) | Time for Addition (10000) | Time for Multiplication (100) |
|---|---|---|
| 12 | 45 ns | 8.98 ms |
| 14 | 46 ns | 187 ms |
| 16 | 49 ns | 337 ms |
| 18 | 78 ns | 1.9 s |
| 20 | 113 ns | 5.4 s |

For the purpose of comparison, we pick the results published in [20], a very practical implementation of BGV scheme. In [20], time taken to compute mean of 100 numbers of size 128-bits is 20 milliseconds, and for variance is 6 seconds. They leave division in both the cases on the data user, and to allow mean computation it requires a 30-bit prime number as secret, while for variance it is 58-bit long. In our implementation, computation of mean takes 1.38 milliseconds and of variance takes 6.83 seconds, including division operation.

## 4.9 Variants

We present here two variants of the scheme.

1. First variant involves larger key size, that is matrices of size 8. This increases computational complexity of the algorithms, but the advantage gained is not much. It obviously increases the ciphertext space thereby contributing to security. In this case encryption algorithm will involve two random numbers. We present here only the encryption algorithm; other algorithms are analogous and can be understood accordingly.

> **Enc$_8$**(x,k)
>
> 1. Choose random values $r_1$ and $r_2$, $r_1, r_2 \in Z_N, r_1, r_2 \neq x$ 2. Construct a matrix $X(m \times 7)$ such that each row has only one element equal to x, three elements equal to r1 and other three equal to r2.
>
> 3. Using Chinese Remainder Theorem, set $x_i, 1 \leq i \leq 7$ to be solution to the simultaneous congruences $x_j = X_{ij} \mod f_i, 1 \leq i \leq m$ .
>
> 4. Ciphertext, $C = k^{-1} * diag(x, x_1, x_2, x_3, x_4, x_5, x_6, x_7) * k$

Here, our aim is to give an idea how the proposed scheme can be generalized to have larger key size, hence better security.

2. Instead of taking a large composite number N as base of ring $Z_N$, it can be chosen as a composite power of 2. Algorithms for all primitives remain exactly the same, except the numbers $p_i$ and $q_i$. They are now selected as powers of 2. All $p_i$ and $q_i$ are unique. For example, for m=2 we can choose p={2, 32} and q={128,8}. Here, the security parameter $\lambda$ can be viewed as the maximum number of bits in a plaintext. This can further be combined with packing bits of plaintext into blocks of $\lambda$ bits each. But it would require an evaluation function which can map binary operations on bits to operations on matrices (or integers).

**Chapter 5**

# CONCLUSION AND FUTURE WORK

Scope and promises of homomorphic cryptography in cloud computing environments cannot be ignored. Researchers all over the world are taking great interest in recent years to develop homomorphisms that can be deployed practically. Much of the focus is on imparting homomorphic capabilities to public key cryptosystems, while some applications can as well be handled with a symmetric key scheme. Hence, our efforts have been to propose ideas as to how symmetric keys and simple matrix-based operations could also lead to feasible schemes for cloud computing, specifically for delegation of computation and private data processing in clouds. communication costs involved in cloud computing are often large, to make up for this we emphasize on having low time complexity for cryptographic primitives.

We have proposed a scheme with a very efficient decryption method hence making it affordable for computationally weak devices, like a mobile device taking results from a computation centre of the cloud and decrypting it. We have proposed application-specific primitives making it easy to deploy to data processing applications. The evaluation functions are efficient and simple making it easy to carry out any arbitrary computation on data. We also suggest how to use symmetric encryption with multiple users, which is clearly key efficient as compared to the popular asymmetric approaches for multiple user applications.

The scheme can be further optimized in matrix multiplication aspect. Decryption need not carry out complete multiplication of three matrices, rather the aim is to derive only the first element of the product matrix.

The scheme can be modified to operate on polynomials instead of working with matrices, deriving idea from [21].

Application to Private information retrieval, searching index of an encrypted database and e-voting can be useful enough. Designing protocols for the same could be a further contribution.

The proposed scheme does not have any scope for targeted malleability or verifiability yet. Improvement in the scheme or introduction of some new primitives for verifiable computation can be appreciable effort.